\documentclass[useAMS]{mn2e}
\usepackage{times}
\usepackage{graphicx}

\title[New {\it HST\/} imaging of six ULX counterparts]{New {\it
Hubble Space Telescope\/} imaging of the counterparts to six
ultraluminous X-ray sources}

\author[T.\,P. Roberts, A.\,J. Levan \& M.\,R. Goad]
{T.\,P. Roberts$^{1}$\thanks{E-mail: t.p.roberts@durham.ac.uk},
A.\,J. Levan$^{2}$ \& M.\,R. Goad$^3$ \\ $^1$ Department of Physics,
Durham University, South Road, Durham DH1 3LE, UK\\ $^2$ Department of
Physics, University of Warwick, Coventry CV4 7AL, UK \\ $^3$ X-ray and
Observational Astronomy Group, Dept. of Physics \& Astronomy,
University of Leicester, University Road, Leicester LE1 7RH, UK
}
\date{}

\def\ro{{\it ROSAT~\/}}

\def\hst{{\it HST~\/}}
\def\chan{{\it Chandra~\/}}

\def\ergsec{{\rm ~erg~s^{-1}}}

\def\H0{{\rm ~km~s^{-1}~Mpc^{-1}}}

\def\cmsq{{\rm ~cm^{-2}}}

\def\la{\mathrel{\hbox{\rlap{\hbox{\lower4pt\hbox{$\sim$}}}{\raise2pt\hbox{$<$}}}}}
\def\ga{\mathrel{\hbox{\rlap{\hbox{\lower4pt\hbox{$\sim$}}}{\raise2pt\hbox{$>$}}}}}

\def\d25{D$_{25}$}

\def\.25{0.25 keV\thinspace}

\begin{document}

\date{Submitted 2006 December 20}

\pagerange{\pageref{firstpage}--\pageref{lastpage}}
\pubyear{2008}

\maketitle

\label{firstpage}

\begin{abstract}
We report the results of new {\it Hubble Space Telescope\/} imaging of
the positions of six ultraluminous X-ray sources (ULXs).  Using images
in three ACS filters we detect good candidate counterparts to four out
of six ULXs, with one more possible detection, and observed magnitudes
in the range $m \sim 22 - 26$ in the F606W filter.  The
extinction-corrected colours and absolute magnitudes vary from source
to source, even after correcting for additional extinction in the host
galaxy, and only one counterpart is readily explained as an OB star.
Nevertheless, these counterparts are decent candidates for future
follow-up in pursuit of dynamical mass constraints on the likely black
holes powering these sources.
%After correcting for extinction within our own Galaxy and the
%host galaxy of the ULX, we derive colours for the counterparts that
%are very blue, with $U - B \sim -2$ and $B - V \sim -0.6$.  We
%conclude that these colours are due to the ULX accretion disc
%dominating the optical emission of the counterparts.  If these ULXs
%are typical of their class, this result implies that obtaining mass
%functions for such objects will be extremely problematic with the
%current generation of telescopes.
\end{abstract}


\section{Introduction}

The ongoing controversy regarding the true nature of the so-called
``ultraluminous X-ray sources'' stems from the simple fact that their
extraordinary observed X-ray luminosities, at $> 10^{39} \ergsec$ in
the 0.5 -- 10 keV band alone, exceeds the Eddington limit for the
stellar mass black holes we observe in our own Galaxy.  An obvious
solution to this problem is that the black holes underlying ULXs are
bigger, perhaps 2 - 3 orders of magnitude more massive
(i.e. intermediate-mass black holes, or IMBHs, with $M_{\rm BH} \sim
10^2 - 10^4 M_{\odot}$, e.g. Colbert \& Mushotzky 1999).  The best
supporting evidence for IMBHs derived from X-ray spectroscopy of
luminous ULXs, that showed possible cool accretion disc spectral
components, having temperatures consistent with discs around $\sim
1000 ~M_{\odot}$ black holes (e.g. Miller, Fabian \& Miller 2004).  In
some cases this argument was supplemented by further evidence, for
example the QPO detections from M82 X-1 (Strohmayer \& Mushotzky
2003).  However, there are strong arguments for the majority of ULXs
possessing stellar-mass black holes, not least the high abundance of
such objects in the most extreme star forming environments (King
2004).  Indeed, detailed re-evaluations of the X-ray spectra of ULXs
have questioned whether the apparent cool accretion discs really
do imply the presence of IMBHs (e.g. Stobbart, Roberts \& Wilms 2006;
Goncalves \& Soria 2006).  If ULXs do not contain the $\sim 1000
~M_{\odot}$ IMBHs suggested by cool discs, then the physical processes
by which smaller black holes (stellar-mass, or perhaps up to $\sim 100
~M_{\odot}$, cf. Stobbart et al. 2006) appear so luminous may include
anisotropic radiation, relativistic beaming and/or truly
super-Eddington emission (e.g. King et al. 2001; K{\"o}rding, Falcke
\& Markoff 2002; Begelman 2002).

%\footnote{The precise dividing line
%between stellar-mass black holes and IMBHs is somewhat indistinct.
%Some authors quote as low as 20 $M_{\odot}$, based on the calculations
%of Fryer \& Kalogera (2001) for individual stellar evolution.
%However, Belczynski et al. (2006) show that black holes in
%mass-transfer binaries can potentially grow up to 100 $M_{\odot}$ in
%size.  Here, we consider the $> 10^2 M_{\odot}$ objects suggested by
%ULX spectroscopy, probably requiring either a primordial origin (Madau
%\& Rees 2001), or to be created in young stellar clusters
%(Portegies-Zwart \& McMillan 2002), as IMBHs.}

Multi-wavelength follow-up - and in particular optical/UV observations
- can provide crucial diagnostics of the nature of ULXs.  Probably the
most important is the identification of an optical counterpart to the
ULX.  The study of such counterparts can reveal the type of the mass
donor star, and subsequently lead to the single most important
measurement for a ULX, namely constraining its mass function (and
hence placing unambiguous limits on the black hole mass) by measuring
a radial velocity curve.  However, even initial identifications are
not trivial in crowded galaxy fields at several Mpc distance.
Fortunately, thanks to a combination of the sub-arcsecond X-ray
astrometry of \chan and the very high spatial resolution and
sensitivity in the optical/UV of the {\it Hubble Space Telescope\/}
({\it HST\/}), good identifications are now possible.  Indeed, many
recent studies have taken advantage of the excellent data from these
missions to identify possible ULX counterparts (e.g. Goad et al. 2002;
Liu, Bregman \& Seitzer 2002, 2004; Soria et al. 2005; Kuntz et
al. 2005; Ramsey et al. 2006; Terashima, Inoue \& Wilson 2006; Ptak et
al. 2006; Mucciarelli et al. 2007; Liu et al. 2007).  Where such
studies have focussed on nearby ($d < 10$ Mpc) spiral galaxies,
individual counterparts with magnitudes $m_{\rm V} \sim 22 - 26$ and
blue colours have frequently been found, consistent with the
interpretation of ULXs as high-mass X-ray binaries, where the mass
donor star is a giant (or supergiant) O4 - B8 star filling its Roche
lobe.  However, an alternative possibility is that the blue colours
are due to reprocessed emission from an X-ray illuminated accretion
disc (Kaaret, Ward \& Zezas 2004; Rappaport, Podsiadlowski \& Pfahl
2005; Copperwheat et al. 2005).  Indeed, Pakull, Gris{\'e} \& Motch
(2006) argue that the blue counterpart to NGC 1313 X-2 is dominated by
accretion disc light (although see Liu et al. 2007 for
counter-arguments).  Optical spectroscopy shows a variation in the
centroid of the He II $\lambda 4686$ emission line from the disc,
which might indicate that an IMBH is ruled out for this source, though
further observations are required to distinguish whether this is a
true radial velocity variation.

Here, we present the results of a programme to produce a deep,
homogeneous set of imaging data for six nearby ULXs, designed to
provide possible targets for future radial velocity studies.  We
selected ULXs from the \ro catalogues of Roberts \& Warwick (2000) and
Colbert \& Ptak (2002) based on the following criteria: (i) distance
$d < 8$ Mpc, to potentially resolve spatial scales below $\sim 2$ pc
using \hst Advanced Camera for Surveys (ACS) data; (ii) the ULX is
located away from the centre of the host galaxy, to avoid inherently
dense stellar environments where individual identifications are
unlikely; and (iii) no \hst imaging data at the position of the ULXs
existed at the time of sample selection.  In addition, the selection
of \ro ULXs (i.e. ULXs detected at relatively soft X-ray energies,
$\leq 2$ keV) ensures a relatively low extinction in the host
galaxies.  The six target ULXs are listed in Tables~\ref{ulxpos} and
2, and we discuss the identification of optical counterparts to these
objects in the following sections.

\section{Data reduction}

Our target ULXs were observed by \hst between September 2005 and
February 2006 (programme number 10579; see Table 2 for details).  A
standard set of observations were obtained for each source,
concentrated primarily in the blue in order to match the colours of
known ULX counterparts.  Hence we obtained a single orbit (four
exposures) of imaging in the F330W filter using the ACS High
Resolution Camera (HRC), and half an orbit (two exposures apiece) in
each of the F435W and F606W filters using the ACS Wide Field Camera
(WFC).  The data was reduced to provide cleaned and stacked images in
each filter via the standard multidrizzle software, with
\textsc{pixfrac} set to one since the data were not well dithered.

\begin{table*}
\centering
\caption{Positions of the target ULXs as obtained from \chan data, and the corresponding Galactic extinction.}
\begin{tabular}{llcccccccc}\hline
ULX	& \chan data$^a$	& RA$^b$	& Dec$^b$	& $\sigma_{\rm cent}~^c$	& $\sigma_{\rm astr}~^d$ 	& \multicolumn{2}{c}{$N_{\rm H, X}~^e$}	& $d~^f$ & $E(B-V)^g$  \\\hline 
IC 342 X-1	& 600254	& 03 45 55.590	& +68 04 55.52	& 0.02	& 0.55 & $\sim 3$ & (1,2)	& 3.9 	& 0.561 \\
IC 342 X-2	& 600273 (H)	& 03 46 15.745	& +68 11 12.59	& 0.10	& 0.61 & $\sim 10$ & (2)	& 3.9  	& 0.558 \\
NGC 2403 X-1	& 600274 (H)	& 07 36 25.525	& +65 35 39.97	& 0.05	& 0.65 & $\sim 3$ & (3)	& 4.2  	& 0.037 \\
NGC 4485 X-1	& 600381	& 12 30 30.487	& +41 41 42.24	& 0.02	& 0.54 & $\sim 3$ & (4)	& 7.8  	& 0.022 \\
%NGC 4485 X-2$^f$& 600381	& 12 30 31.647	& +41 41 40.95	& 0.03	& 0.54 \\
NGC 5055 X-2	& 700387	& 13 15 19.578	& +42 03 01.92	& 0.05	& 0.59 	& $\sim 1$	& (5) 	& 7.2  	& 0.018 \\
M83 IXO 82	& 600477 (H)	& 13 37 19.801	& -29 53 48.72	& 0.02	& 0.55 & $\sim 1$ & (3)	& 4.7  	& 0.063 \\
\hline
\end{tabular}
\begin{minipage}[t]{6.6in}
Notes: $^a$ \chan observation sequence number.  An appended '(H)'
indicates that the observation was with the HRC-I detector; otherwise
observations were taken with the ACIS-S.  $^b$ Right ascension and
declination (epoch J2000) in hexagesimal format.  $^c$ Error on
position of the centroid of the X-ray source, in arcseconds.  $^d$
Error on the astrometric precision of the X-ray source position in
arcseconds. $^e$ Absorption column towards the ULX, external to our
own Galaxy ($10^{21} \cmsq$), measured by X-ray spectroscopy.
Measurements from (1) Roberts et al. (2004); (2) Bauer, Brandt \&
Lehmer (2003); (3) Stobbart, Roberts \& Wilms (2006); (4) Roberts et
al. (2002); (5) analysis of the \chan data, using standard reduction
and analysis techniques, and an absorbed power-law continuum model
(giving $\Gamma \sim 2.5$).  $^f$ Distance to host galaxy in Mpc, from
papers quoted in the previous column.  $^g$ Line-of-sight foreground
Galactic extinction for each ULX counterpart, inferred from Schlegel,
Finkbeiner \& Davis (1998).
\end{minipage}
\label{ulxpos}
\end{table*}

In five of the six cases we obtained X-ray positions from archival
\chan data.  In the sixth case (M83 IXO 82) we were awarded a new
\chan High Resolution Camera for Imaging (HRC-I) observation to obtain
an accurate position, as the ULX lay outside the field-of-view of
previous \chan observations of that galaxy.  Archival data from both
the HRC-I and ACIS (either -S or -I) were considered, and that with
the best combination of exposure length and proximity of the ULX to an
on-axis position selected.  These data sets are listed in
Table~\ref{ulxpos}.  The astrometry of the archival data was initially
corrected for known aspect offsets\footnote{See {\tt
http://cxc.harvard.edu/cal/ASPECT/fix\_offset}.}, before the data were
reprocessed using \textsc{ciao} version 3.3.0.1 and the \chan
calibration database version 3.2.1.  Broad band (0.5 -- 8 keV) images
were extracted from the reprocessed event files, and searched for
sources using the \textsc{wavdetect} algorithm.  We list the derived
ULX positions and centroiding errors ($\sigma_{\rm cent}$) in
Table~\ref{ulxpos}.

%Note that this Table includes a
%seventh ULX, NGC 4485 X-2, located $\sim 10$ arcseconds to the east of
%NGC 4485 X-1.  This is a new source, that turned on between the
%original \chan observations of NGC 4485/90 (in November 2000; Roberts
%et al. 2002) and subsequent follow-ups (July/November 2004).  An
%initial examination of its characteristics suggest a luminosity of
%$\sim 10^{39} \ergsec$, hence as its position is covered by our \hst
%imaging we include it here as a candidate ULX.

\section{Identification and characterisation of the possible new counterparts}

The key step in identifying optical counterparts to our target ULXs is
registering the relative astrometries of the \hst and \chan data.
Ideally this would be done by finding a number of \hst counterparts to
\chan X-ray sources; however there were insufficient matches present
in our data.  Instead we aligned our ACS images to 2MASS (which
typically had the largest number of sources in an ACS field), or, in
the case of NGC 4485 where only two 2MASS sources fell within the ACS
field, we performed relative astrometry using observations of NGC 4485
obtained at the Isaac Newton Telescope on 2001 January 4.  We then
took advantage of the small intrinsic errors in \chan astrometry (90
per cent confidence region of 0.5 arcseconds on the absolute position
accuracy, after known aspect errors are removed, for both the HRC-I
and ACIS-S\footnote{See {\tt
http://cxc.harvard.edu/cal/ASPECT/celmon/}.}) and overlaid error
regions on the corrected \hst images with radii that were a
combination (in quadrature) of this \chan uncertainty and the residual
uncertainty in the \hst astrometry after correction to 2MASS.  The
resulting error regions - typically $\sim 0.6$ arcseconds radius - are
shown in Table~\ref{ulxpos}.

\begin{figure*}
\centering
\includegraphics[width=5.6cm]{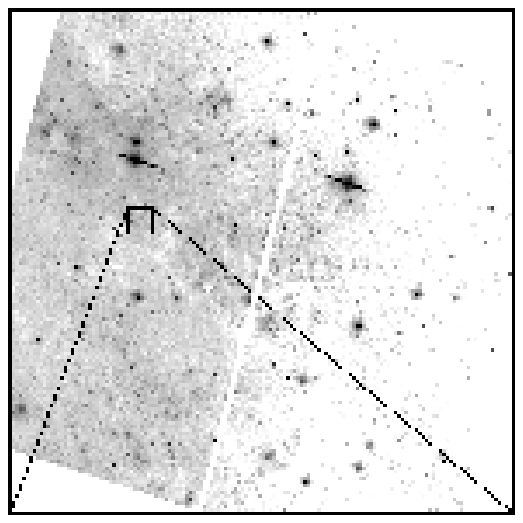}
\includegraphics[width=5.6cm]{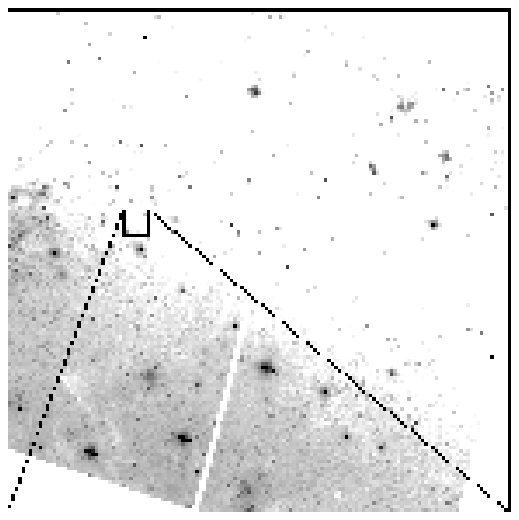}
\includegraphics[width=5.6cm]{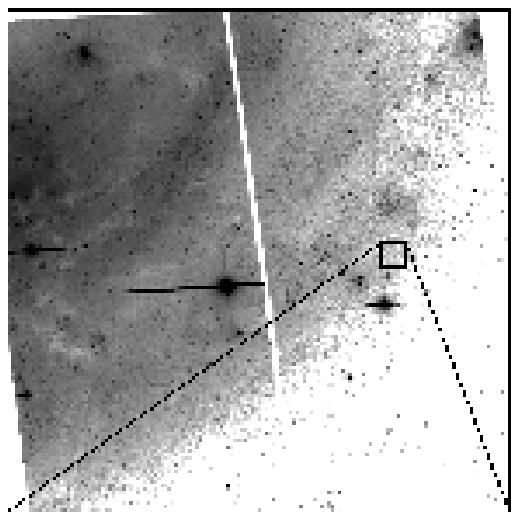}\\
\includegraphics[width=5.6cm]{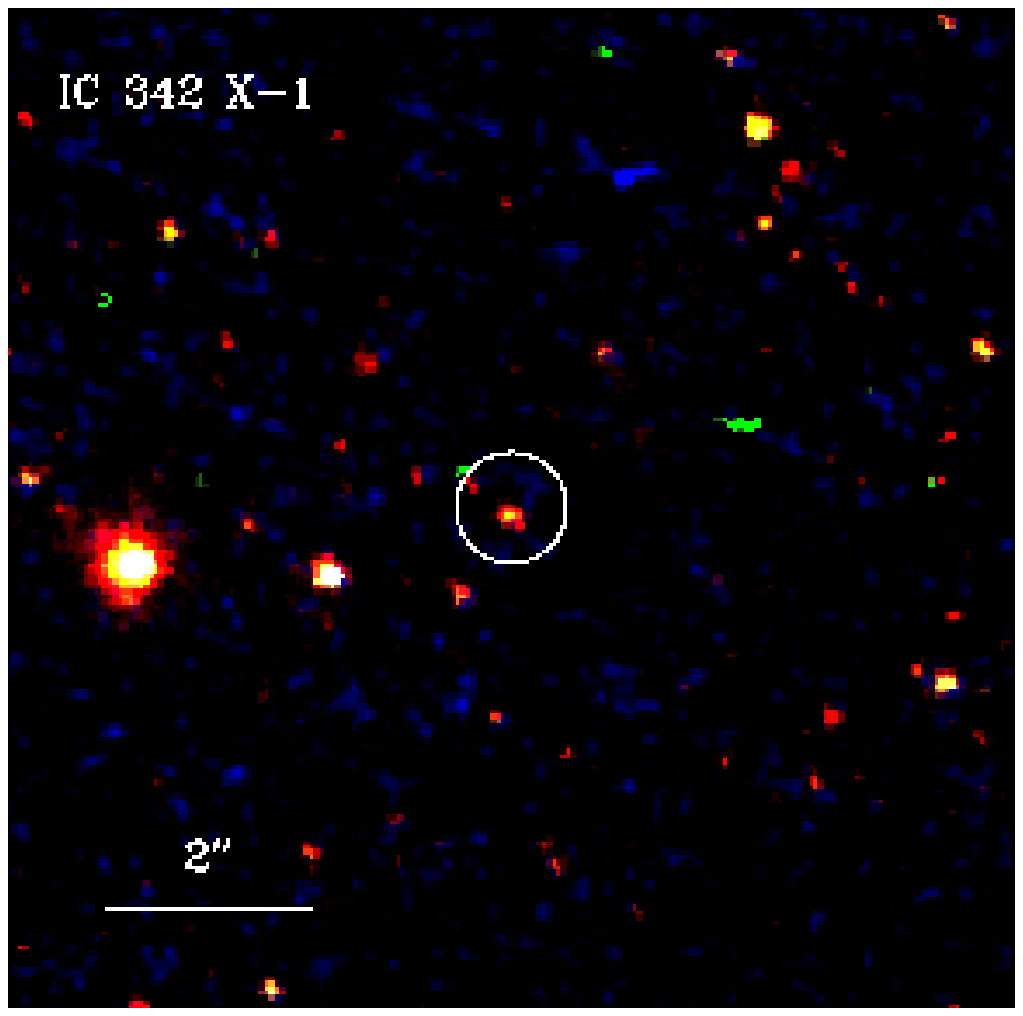}
\includegraphics[width=5.6cm]{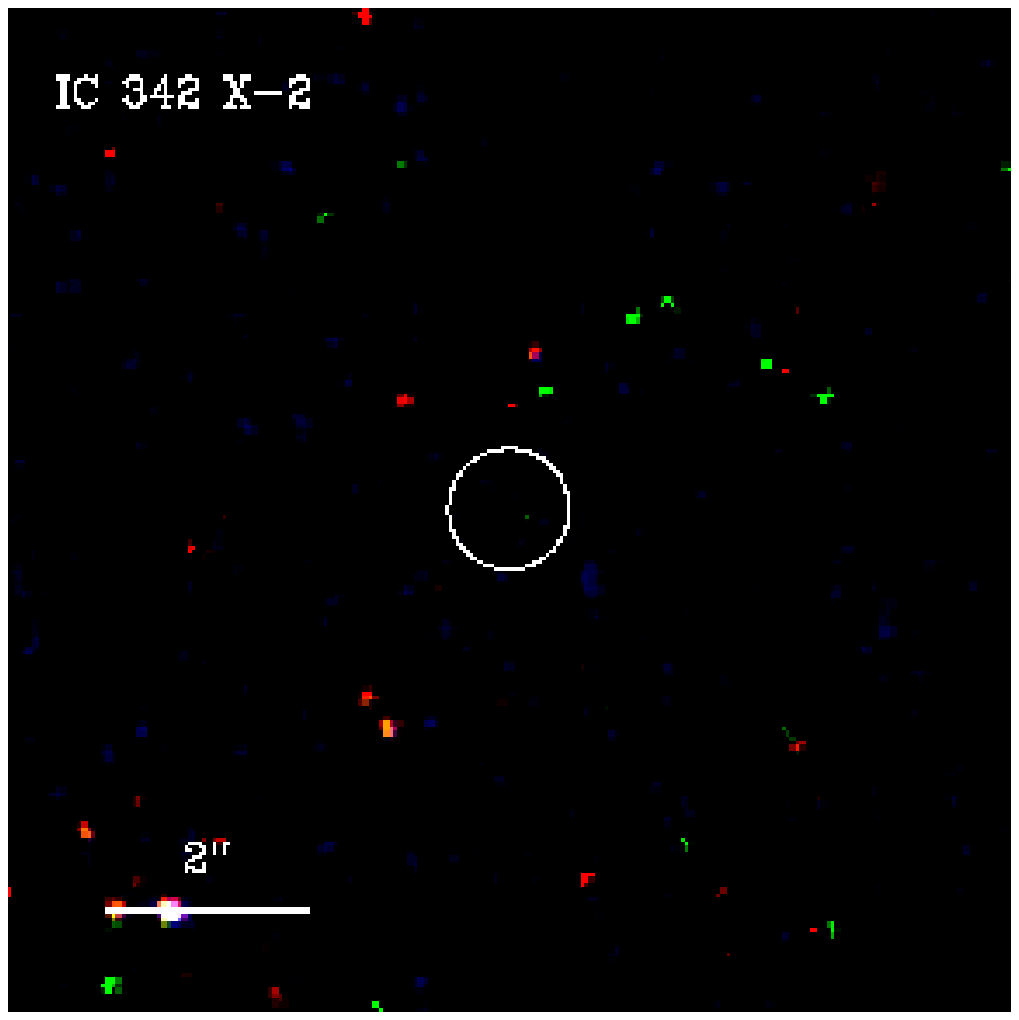}
\includegraphics[width=5.6cm]{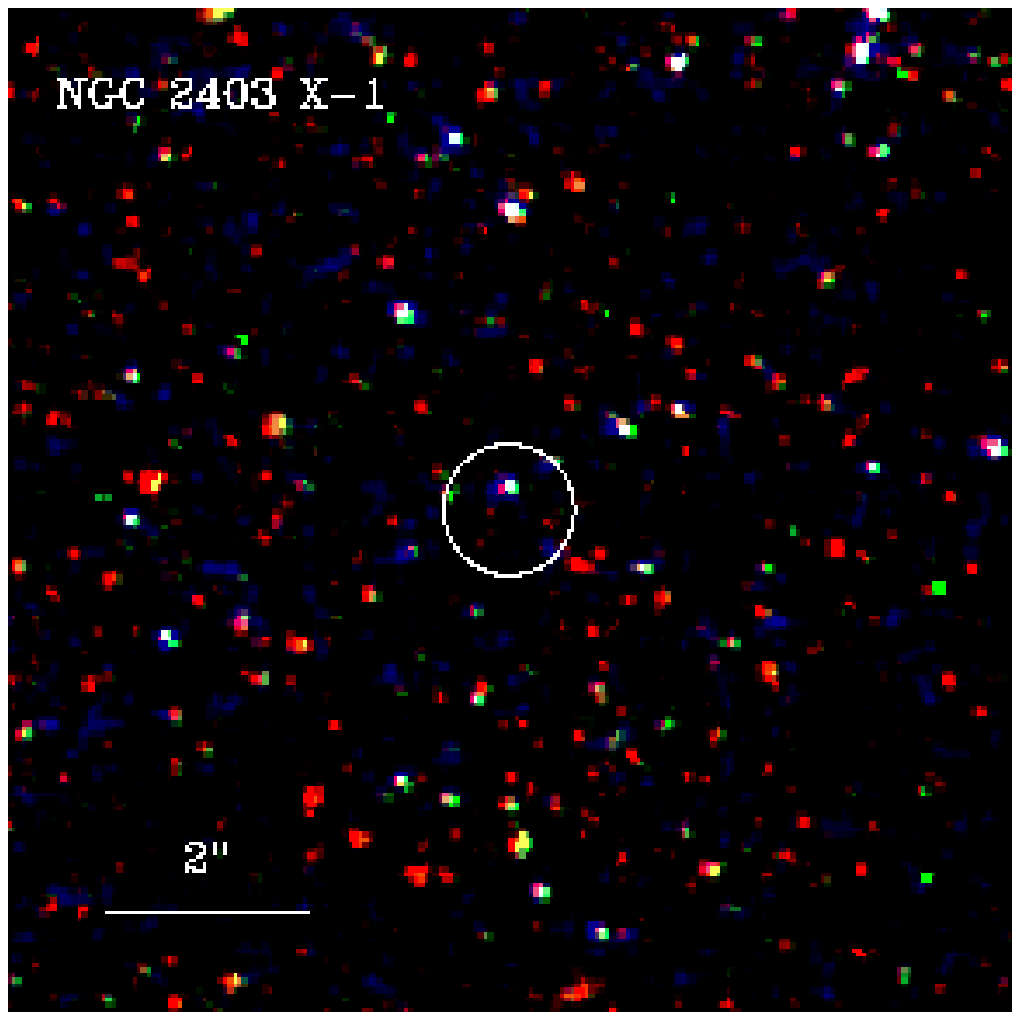}
\includegraphics[width=5.6cm]{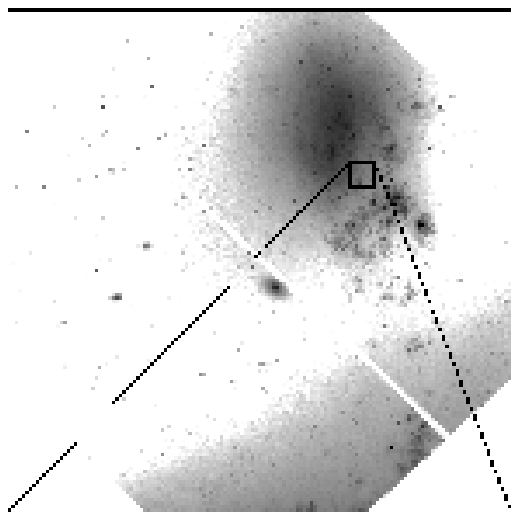}
\includegraphics[width=5.6cm]{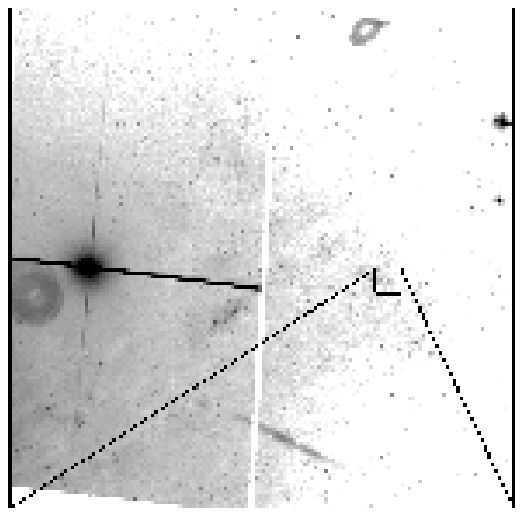}
\includegraphics[width=5.6cm]{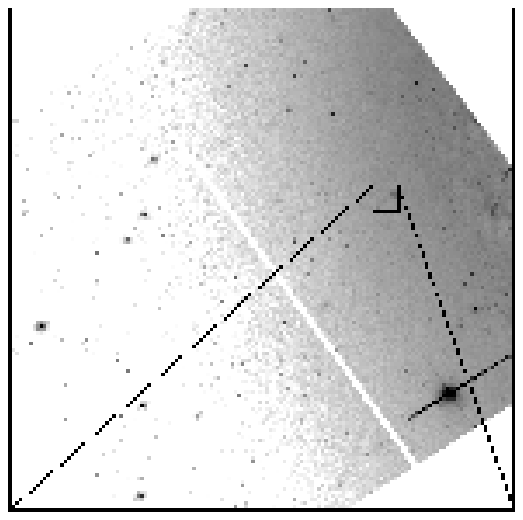}\\
\includegraphics[width=5.6cm]{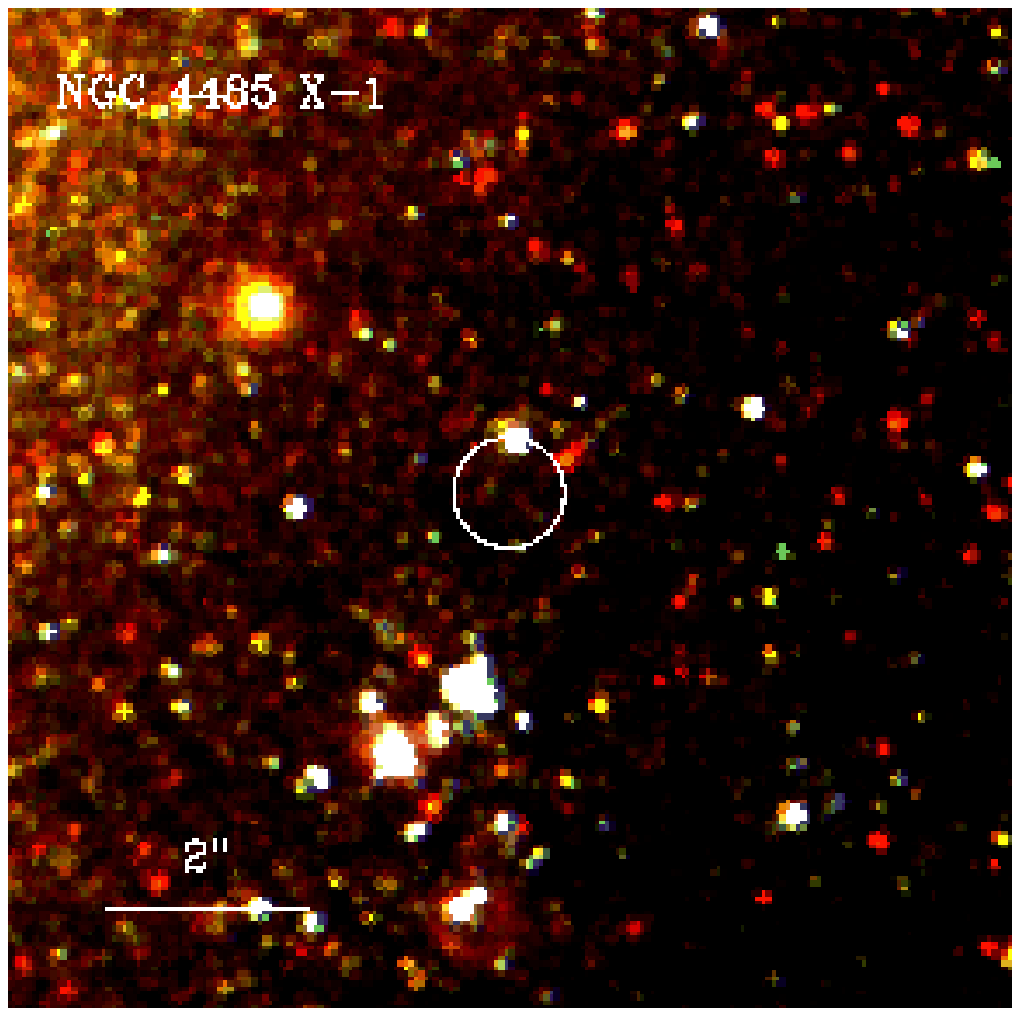}
\includegraphics[width=5.6cm]{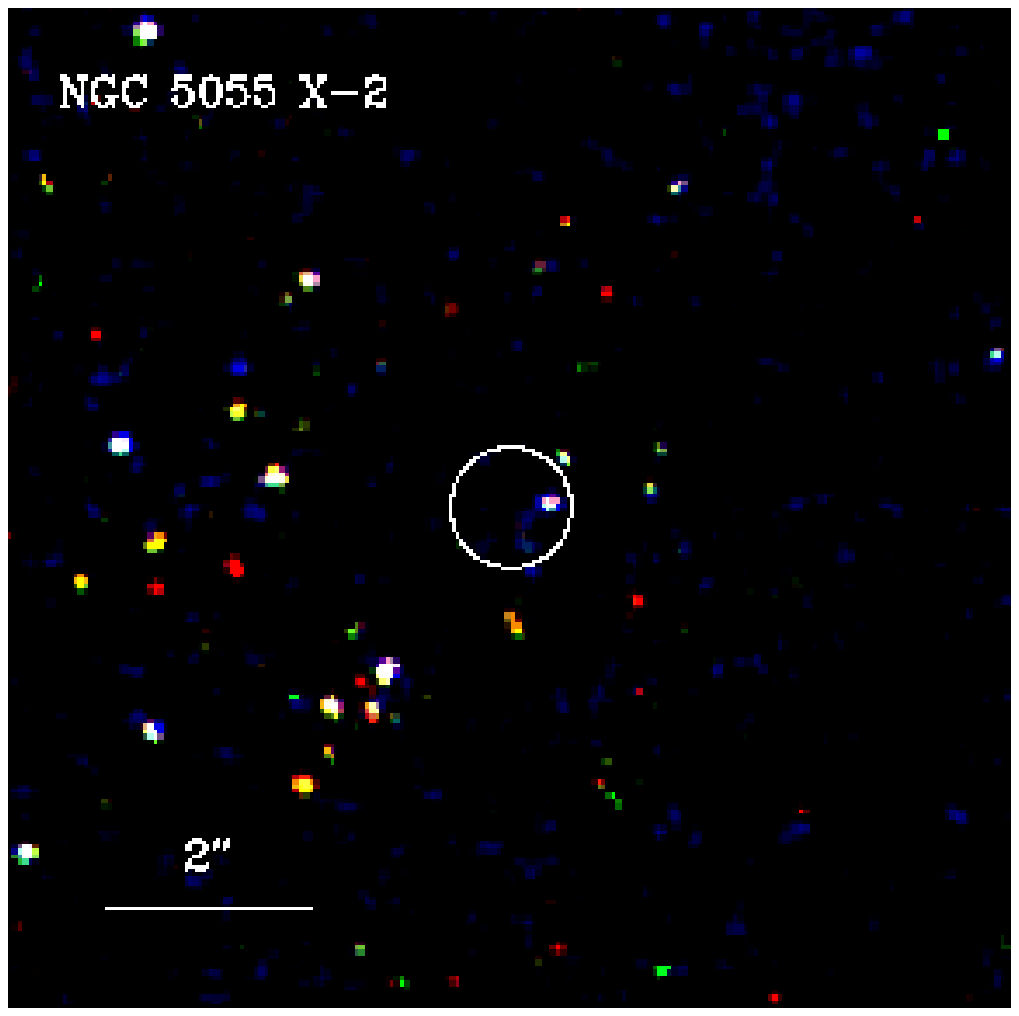}
\includegraphics[width=5.6cm]{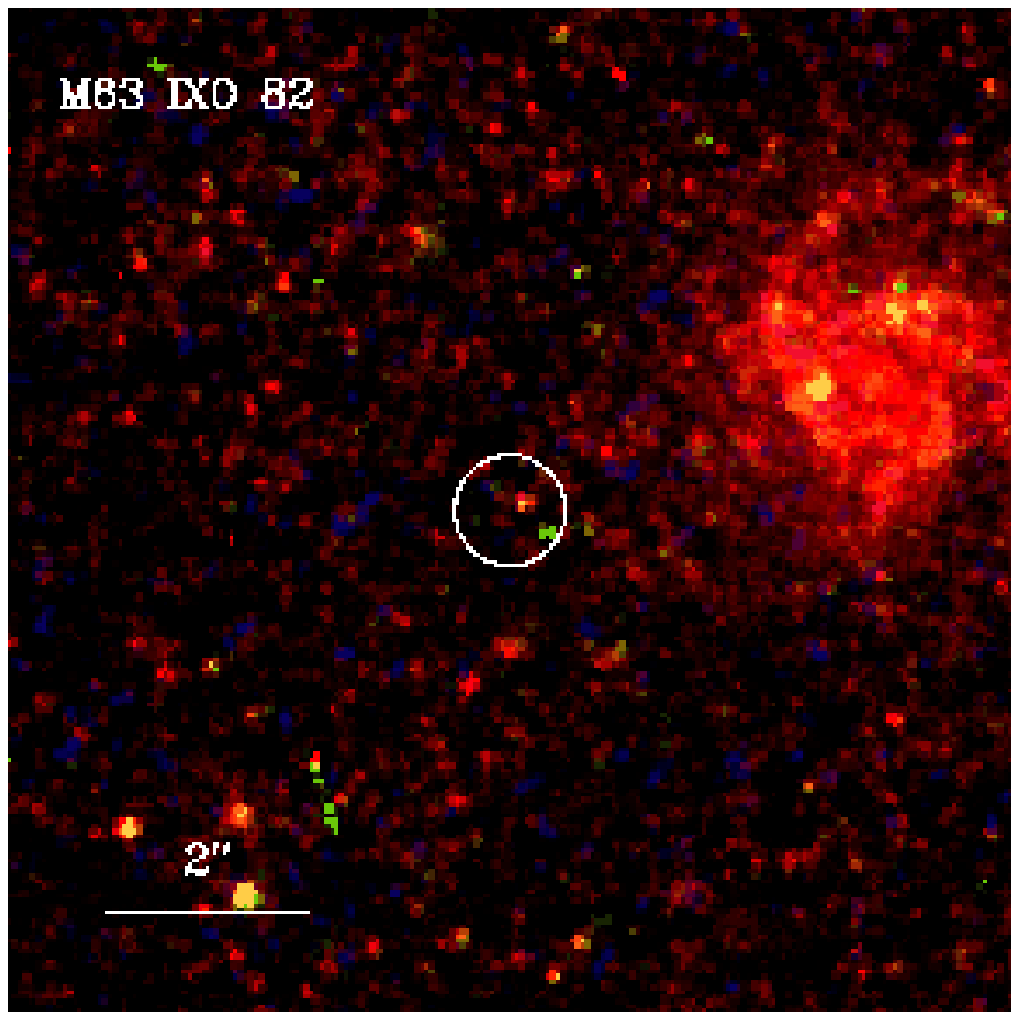}
\caption{\hst ACS images of ULX locations.  North is up in each
image.  For each of the six targets, we show a wide ACS/WFC F606W view
of the ULX environment, with a true-colour blow-up of the immediate
vicinity of the ULX beneath it (Blue = F330W, Green = F435W \& Red =
F606W).  The error circles are as per Table~\ref{ulxpos}.}
\label{hstimgs}
\end{figure*}

The locations of the ULXs are shown in Figure~\ref{hstimgs}.  In
several cases a relatively bright counterpart is clear.  In other
cases multiple counterparts are present, or (in the case of IC 342
X-2) a counterpart is indiscernable.  We take the view that the most
likely counterpart to the ULX is the brightest source within the error
circle or, in the case of multiple sources, that with the bluest
colours.  We tabulate the observed and inferred characteristics of
these candidates in Table 2.  The magnitudes of the counterparts were
derived using the following procedure.  Firstly, magnitudes were
derived at the pivot wavelength of each ACS filter using a small (0.1
arcsecond) aperture, with zeropoints taken from Sirianni et
al. (2005), and additional aperture corrections calculated using
\textsc{synphot}.  This is shown as Mag (pivot) in Table 2.  These
magnitudes were then corrected for reddening within our own Galaxy
using the value of $E(B-V)$ in Table 1 and the filter-specific
extinction ratios given by Sirianni et al. (2005) for an O5 V star.
Finally, each filter magnitude was photometered into the appropriate
tabulated $UBV$ magnitude (Vega mag in Table 2, with F330W converting
to $U$, etc) using a power-law continuum slope derived from the
relative extinction-corrected magnitudes in the ACS filters as a proxy
for the spectral shape.  In the few cases where we saw no counterpart,
we seeded the appropriate images with fake stars in order to infer a
detection limit at the position of the ULX.  These limits are also
shown in Table 2.

The extinction-corrected and photometered magnitudes were converted to
absolute magnitudes using the distances to the host galaxies in
Table~\ref{ulxpos}, and also used in the calculation of the
counterpart colours $U - B$ and $B - V$.  However, these values do not
allow for any extinction within the host galaxy of the ULX.  As the
absorption column imprinted on the X-ray spectra of the ULXs appears a
good measure of the column through the host to the ULX position in
most cases (Winter, Mushotzky \& Reynolds 2007), we use the values in
Table~\ref{ulxpos} to infer an additional, host galaxy V-band
extinction by considering a column of $10^{21} \cmsq$ to correspond to
$A_{\rm V} = 0.56$ magnitudes of extinction (Predehl \& Schmitt 1995).
This additional extinction was also applied to the ACS magnitudes,
before photometering by the method described above.  The results of
applying this second correction are shown in parentheses for the
absolute magnitudes and colours of the counterparts in Table 2.

\setcounter{table}{1}
\begin{table*}
\begin{center}
\caption{\hst observations of the ULX locations, and characteristics
of the candidate counterparts.}
\begin{tabular}{lcccccccc}\hline
ULX & Filter & Observation date & Exposure & Mag (pivot)$^a$ & Vega mag$^b$ & $M_{\rm V} ~^c$ & $U - B~^d$ & $B - V~^d$ \\
& & & (s) & \\
\hline
 & F330W &  2005-09-02 03:14:42 & 2900 & $>$25.2 & $U > 22.4 $ &  \\
IC342 X-1 & F435W & 2005-09-02 01:19:59 & 1248 & 25.21 $\pm$ 0.07 & $B = 22.98 \pm 0.07$ & -5.57 (-7.30) & $>$-0.6 ($>$-1.1) & 0.60 (0.16)\\\vspace*{0.15cm}
 & F606W  & 2005-09-02 01:47:02 & 1248 & 23.91 $\pm$ 0.05 & $V = 22.38 \pm 0.06$ & \\
%\hline
& F330W  & 2005-09-02 06:31:42 & 2900 & - & - \\
IC342 X-2 & F435W  & 2005-09-02 06:31:42 & 1248 & 26.71 $\pm$ 0.26 & $B \approx 24.5$ &  - & - & - \\\vspace*{0.15cm}
& F606W  & 2005-09-02 04:58:42 & 1248 & -  & - \\
%\hline
& F330W  & 2005-10-17 04:22:04 & 2912 & 24.82 $\pm$ 0.14 & $U = 23.87 \pm 0.14$ & \\
NGC 2403 X-1 & F435W & 2005-10-17 02:23:25  & 1248 & 24.74 $\pm$ 0.11 & $B = 24.74 \pm 0.11$ & -3.42 (-5.14) & -0.87 (-1.20) & 0.04 (-0.42) \\\vspace*{0.15cm}
& F606W & 2005-10-17 02:50:28  & 1248 & 24.90 $\pm$ 0.11 & $V = 24.70 \pm 0.11$ & \\
%\hline
& F330W &  2005-11-19 12:42:30 &  2652 &22.20 $\pm$ 0.04 & $U = 21.31 \pm 0.04$ & \\
NGC 4485 X-1 & F435W  & 2005-11-19 07:56:30 & 1116 &21.73 $\pm$ 0.03 & $B = 21.78 \pm 0.03$ & -7.56 (-9.28) & -0.47 (-0.82) & -0.12 (-0.59) \\ \vspace*{0.15cm}
& F606W   & 2005-11-19 09:18:42 & 1116 & 22.02 $\pm$ 0.03 & $V = 21.90 \pm 0.04$ & \\
%\hline
%& F330W &  2005-11-19 12:42:30 & confused & & 0.12\\
%NGC 4485 X-2 & F435W  & 2005-11-19 07:56:30 & confused & & 0.09 & - & - \\ 
%& F606W   & 2005-11-19 09:18:42&confused & & 0.06\\
%\hline
& F330W & 2006-01-29 20:13:18 & 2652 & 24.15 $\pm$ 0.06 & $U = 23.35 \pm 0.06$ & \\
NGC 5055 X-2 & F435W & 2006-01-22 02:41:42 & 1116 &24.71 $\pm$ 0.04 & $B = 24.79 \pm 0.04$ & -4.64 (-5.35) & -1.44 (-1.46) & 0.14 (0.14) \\\vspace*{0.15cm}
& F606W & 2006-01-22 04:15:29 & 1116 & 24.85 $\pm$ 0.03 & $V = 24.65 \pm 0.04$ & \\
%\hline
& F330W & 2006-02-25 19:43:52  & 2568 & $>$25.0 & $U > 24.7$ \\
M83 IXO 82 & F435W & 2006-02-25 17:46:39  & 1000 & 25.79 $\pm$ 0.13 & $B = 25.66 \pm 0.13$ & -3.00 (-3.57) & $>$-1.0 ($>$-1.1) & 0.30 (0.15) \\
& F606W & 2006-02-25 18:09:34  & 1000 & 25.57 $\pm$ 0.16 & $V = 25.36 \pm 0.17$ & \\
\hline
\end{tabular}
\begin{minipage}[t]{6.8in}
Notes: $^a$ Magnitude of the counterpart at the pivot wavelength of
the ACS filter.  The uncertainty in the measurement is calculated by
seeding the image with fake stars of the same magnitude.  The two
limits are calculated by seeding the images with fake stars covering a
range of input magnitudes.  $^b$ $UBV$ magnitudes in the Vega
magnitude system.  These are extinction-corrected for material in our
own Galaxy, then directly photometered from the ACS filters as
discussed in the text.  The error includes an additional (relatively
small) uncertainty from the calculated error in the slope of the
power-law continuum used to photometer the data.  We use a $\nu^{-1}$
spectrum to provide the magnitude for the possible IC 342 X-2
counterpart, but quote no limits on non-detections as we cannot obtain
a proxy for the spectral shape of this source from the one data point.
$^c$ Absolute magnitude of the counterpart at the distance of its host
galaxy (also foreground extinction corrected).  Figures in parentheses
include a further correction for the excess column to the ULX measured
by X-ray spectroscopy (see text).  $^d$ Foreground extinction
corrected counterpart colours, again shown with a further possible
extinction correction in parentheses.
\end{minipage}
\end{center}
\label{cptfxs}
\end{table*}

\section{The counterparts}

Our observations detect four likely counterparts to ULXs, plus two
more possibles.  Three of these likely counterparts are previously
unreported; the counterpart to IC 342 X-1 was been reported twice
before.  It was described as ``candidate 1'' (of two possible
counterparts) by Gris{\'e}, Pakull \& Motch (2006) based on
ground-based imaging, and latterly identified as the unique
counterpart ``Star A'' in a separate \hst programme by Feng \& Kaaret
(2008).  Although their observations took place 3 months after ours,
Feng \& Kaaret's Star A has identical $B$ and $V$ magnitudes to our
detection (within the quoted errors), suggesting it displays little
intrinsic variability.  From their characterisation of Star A they
derive a classification as an F8 to G0 Ib supergiant, although they
choose to omit any corrections to their characterisation based on
extinction intrinsic to IC 342.  Candidate 1 of Gris{\'e} et
al. (2006) also displays a similar magnitude after extinction
correction ($m_{\rm V} \sim 22.3$).

The likely counterparts are all luminous point-like sources, with
absolute magnitudes in the range $M_{\rm V} \sim -3.4$ to $-7.6$, or
perhaps even higher ($M_{\rm V} \sim -5.1$ to $-9.3$) if we take into
account possible additional extinction in the host galaxies of the
ULXs\footnote{These columns are approximations based on values taken
from a range of models for the X-ray spectra of the ULXs.  As the
variance in measured column between models can be factors at least
$\sim 2$, the uncertainties on the quoted values are large, though we
note that this correction will always work in the sense of increasing
the intrinsic brightness of the counterpart.}.  The inferred $U - B$
colours of these ULXs are very blue, in the range $-1.4$ to $-0.5$, or
$-1.5$ to $-0.8$ accounting for the putative host galaxy reddening.
The $B - V$ colour ranges from $-0.1$ to $0.6$ ($-0.6$ to $0.2$ with
host galaxy reddening).

Of the two possible counterpart detections, IC 342 X-2 is very
marginal, with a minimal detection in the F435W filter image alone.
This is perhaps not surprising, given the high foreground Galactic
extinction to IC 342, and the even higher X-ray column, which may
result in $> 5$ magnitudes of extinction in the V-band alone.  In this
regard, the detection in the F435W filter but not the F606W is very
suspicious, and increases the chances of this being a spurious
detection.  M83 IXO 82, on the other hand, probably is a real
detection.  The question here is whether it is the correct one.  It is
suspiciously close (at $\sim 5$ arcseconds distance) from the centre
of a galaxy lying behind a spiral arm of M83, which does raise the
question of whether this \chan HRC data set has an unusually bad
aspect solution, and the X-ray source is in fact located within the
background galaxy (probably an AGN).  However, to argue against this,
the colours of the counterpart are somewhat similar to IC342 X-1, and
(though it is fainter than IC342 X-1 by $\sim 2.5$ magnitudes) it is
only $\sim 1.5$ magnitudes intrinsically fainter then NGC 2403 X-1.
Clearly future work should attempt to clarify the astrometry for this
field and confirm the likely nature of this X-ray source.

Previous observations have also revealed blue counterparts to ULXs,
that have generally been identified as O or B stars.  In order to
clarify whether we are viewing similar objects, we referred to the
work of Wegner (2006) to provide us with typical $M_{\rm V}$
measurements for such stars, and Martins \& Plez (2006) and Knyazeva
\& Kharitonov (1998) for typical colours.  These colours for OB stars
cover ranges of $U-B \sim -1.2 \rightarrow -0.6$ and $B-V \sim -0.3
\rightarrow -0.1$.  Hence, taking only the Galactic-extinction
corrected magnitudes into consideration, the four good candidate
counterparts in general appear too red to be OB stars.  This situation
is however mostly remedied when we take into account possible
extinction in the host galaxy - in most cases the colours
(particularly the $U-B$ colour) become sufficiently blue to originate
in OB stars.  However, the inherent uncertainty in this additional
correction is large (see footnote 3), so it is perhaps sensible to
simply conclude that the Galactic-extinction corrected magnitudes
provide an upper (reddened) limit on the colours, which are
intrinsically bluer than this value.  Indeed, if some fraction of the
additional X-ray absorption is local to the ULX accretion disc, or if
the intense X-ray emission of the ULX can deplete the dust content of
its local environment, it is quite plausible that the extinction to
the optical counterpart within the host galaxy could be lower than
estimated.  Evidence supporting this is provided by Gris{\'e} et
al. (2006), who infer $E(B-V) = 0.26$ within IC 342 from the
H$\alpha$/H$\beta$ ratio (albeit from the bubble nebula surrounding
this ULX) -- this is somewhat smaller than the value $E(B-V) = 0.53$
within IC 342 derived from the X-ray absorption.

Supposing that the true counterpart colours lie between the tabulated
colours and magnitudes, (i.e. with and without additional host galaxy
extinction), we can speculate on the nature of the counterparts.  The
colours and magnitude of the counterpart to NGC 2403 X-1 suggest it
could plausibly be an O-star or an early B giant/B supergiant.  NGC
4485 X-1 is somewhat trickier to classify - it is relatively blue in
$B-V$ but red in $U-B$ compared to other candidates.  Given its high
absolute magnitude perhaps the most likely explanation is that this
object is a young, compact stellar cluster that has lost its youngest
and hottest OB stars (perhaps similar to HST-1 identified by Goad et
al. 2002 as a possible counterpart to NGC 5204 X-1, though this was
later ruled out by Liu et al. 2004).  The counterpart to NGC 5055 X-2
is an oddball - its absolute magnitude is bright enough for an OB
giant/supergiant, but its $U-B$ colour is too blue, whereas its $B-V$
colour is too red.  It is unclear what is causing this unusual colour,
though domination of the optical light by the accretion disc can be
ruled out, as this will produce a spectrum redder than an O-star.  We
speculate that this blueness could be caused by emission lines and/or
non-stellar processes, and may possibly even be attributable to
variability in the accretion disc, however this cannot be confirmed
without repeated spectroscopic observations.  Finally, the $B-V$
colour of the counterpart to IC 342 X-1 is quite red, even after
correction, apparently ruling out an OB star unless substantial
additional local extinction (not seen in X-rays) is present.  In fact,
its red colour and magnitude are more reminiscent of a F-type
supergiant, probably earlier in type (F0 - F5 supergiant) than
suggested in Feng \& Kaaret (2008) after accounting for the additional
column within IC 342.  Again, this source would benefit from further
deep spectroscopic follow-up to clarify its nature.

Unlike previous studies, it appears that few of our counterparts can
be easily associated with individual OB stars.  This may be a result
of detecting fainter targets than previously observed (and hence
poorer data quality), a situation that could be improved with future,
deeper observations.  Indeed, a combination of deeper observations and
the new generation of models for the optical emission of ULXs, such as
those of Copperwheat et al. (2007), Patruno \& Zampieri (2008) and
Madhusudhan et al. (2008), should reveal more of the nature of ULXs.
So far the work of both Copperwheat et al. and Patruno \& Zampieri has
suggested that the currently-known counterparts of ULXs are older and
less massive than initially suggested (due to the modifying effects
of, for example, the illumination of the donor star by the ULX on the
optical emission of these systems).  All authors agree that the
presence of an IMBH will increase the optical luminosity of the
systems, and indeed Madhusudhan et al. claim that the colours of
current ULX counterparts appear more consistent with their models
incorporating the presence of an IMBH.

%Nonetheless, our detected counterparts are
%generally blue, and so the question of whether their emission is
%dominated by stellar light or the reprocessed emission from the
%accretion disc is an interesting one, not least because it impacts on
%their suitability for future optical spectroscopic follow-up, with a
%view to determining radial velocity curves and hence mass functions.
%The work of Copperwheat et al. (2007) is important in this respect.
%They fit models of the optical light from an X-ray irradiated
%accretion disc and secondary star system to ULX counterpart data, that
%suggest that these counterparts are generally older and less massive
%than previously reported.  The previous underestimation of their age
%and mass is primarily due to heating of the secondary star by the
%accretion disc; however, the counterparts must still be B stars.  This
%is encouraging for follow-up work, as it suggests that the optical
%spectra of ULX counterparts is dominated by stellar
%light\footnote{Also see the recent papers by Patruno \& Zampieri
%(2008) and Madhusudhan et al. (2008) for further modelling and
%discussion of the optical emission of ULX donor stars and accretion
%discs.}.  

However, the most important observations remain those that could
constrain the mass of a ULX dynamically.  Rather encouragingly, recent
work by Orosz et al. (2007), Prestwich et al. (2007) and Silverman \&
Filippenko (2008) show that this can be done for massive stellar
donors in X-ray binary systems out to distances $\sim 1$ Mpc.  To
perform similar observations - particularly based on the He {\small
II} $\lambda 4686$ line used by Prestwich et al. (2007) and Silverman
\& Filippenko (2008) - on the slightly more distant ULX counterparts
may therefore only be a small step away.  The counterparts revealed by
this study are promising candidates for such observations, albeit
technically challenging ones for the immediate future due to their
relatively faint magnitudes.  Such studies should be pursued
regardless, as the prize - a dynamical mass constraint on the black
hole in an ULX - is the single most important measurement to be made
in this field.

\vspace{0.2cm}
{\noindent \bf ACKNOWLEDGMENTS}\\ The authors thank the referee
(Manfred Pakull) for comments that have substantially improved this
paper.  We also gratefully acknowledge PPARC for financial support
during the early stages of this work.  Based on observations made with
the NASA/ESA Hubble Space Telescope, obtained at the Space Telescope
Science Institute, which is operated by the Association of
Universities for Research in Astronomy, Inc., under NASA contract NAS
5-26555. These observations are associated with program 10579.

%TPR gratefully acknowledges support
%from the UK Particle Physics and Astronomy Research Council (PPARC).
%The authors would also like to thank Simon Vaughan for his advice on
%temporal analyses, and Chris Done for discussions on Galactic black
%holes.
%This work is based on observations obtained with
%{\it XMM-Newton\/}, an ESA science mission with instruments and
%contributions directly funded by ESA member states and the USA (NASA).

\label{lastpage}

\end{document}